\documentclass[%
reprint,
superscriptaddress,
amsmath,amssymb,
aps,
prl,
]{revtex4-1}

\usepackage{graphicx}
\usepackage{dcolumn}
\usepackage{bm}
\usepackage{xcolor}

\begin{document}


\title{Tracking dark excitons with exciton-polaritons in semiconductor microcavities}
\author{D.~Schmidt}
\author{B.~Berger}
\author{M.~Kahlert}
\affiliation{
	Experimentelle Physik 2,
	Technische Universit\"at Dortmund,
	D-44221 Dortmund, Germany
}  
\author{M.~Bayer}
\affiliation{
	Experimentelle Physik 2,
	Technische Universit\"at Dortmund,
	D-44221 Dortmund, Germany
}
\affiliation{
	A. F. Ioffe Physical-Technical Institute,
	Russian Academy of Sciences,
	St Petersburg 194021, Russia
}
\author{C.~Schneider}
\affiliation{
	Technische Physik,
	Universit\"at W\"urzburg,
	97074 W\"urzburg, Germany
}
\author{S.~H\"ofling}
\affiliation{
	Technische Physik,
	Universit\"at W\"urzburg,
	97074 W\"urzburg, Germany
}

\affiliation{
 SUPA, School of Physics and Astronomy, 
University of St. Andrews, 
St. Andrews KY16 9SS, UK
}

\author{E.~S.~Sedov}
\affiliation{
	School of Physics and Astronomy,
	University of Southampton,
	SO17 1NJ Southampton, United Kingdom
}
\affiliation{
	Vladimir State University named after A. G. and N. G. Stoletovs,
	Gorky str. 87, 600000, Vladimir, Russia
}

\author{A.~V.~Kavokin}
\affiliation{
	Spin Optics Laboratory,
	St. Petersburg State University,
	Ul’anovskaya 1, Peterhof, St. Petersburg 198504, Russia
}
\affiliation{
	International Center for Polaritonics,
	Westlake University,
	No.~18, Shilongshan Road, Cloud Town, Xihu District, Hangzhou, China
}

\author{M.~A\ss mann}
\affiliation{
	Experimentelle Physik 2,
	Technische Universit\"at Dortmund,
	D-44221 Dortmund, Germany
}

\date{\today}
\begin{abstract}
{Dark excitons are of fundamental importance for a wide variety of processes in semiconductors, but are difficult to investigate using optical techniques due to their weak interaction with light fields. We reveal and characterize dark excitons non-resonantly injected into a semiconductor microcavity structure containing InGaAs/GaAs quantum wells by a gated train of eight 100\,fs-pulses separated by 13\,ns by monitoring their interactions with the bright lower polariton mode. We find a surprisingly long dark exciton lifetime of more than 20\,ns which is longer than the time delay between two consecutive pulses. This creates a memory effect that we clearly observe through the variation of the time-resolved transmission signal. We propose a rate equation model that provides a quantitative agreement with the experimental data.}
\end{abstract}

\pacs{Valid PACS appear here}
\maketitle
A detailed understanding of the nature of electronic excitations in semiconductor crystals is fundamental in order to explain their dynamics, collective interactions and many-body effects. Optical spectroscopy provides a convenient range of characterisation tools for those excitations that are bright, which means that they can absorb or emit light. It is much more complicated to gain experimental access to optically inactive or dark excitations which interact weakly or not at all with light. So called dark excitons are typical representatives of such excitations.
Still, their properties are decisive for a wide range of systems ranging from semiconductor monolayers~\cite{Malic2018,Lundt2017,Wang2017} and light harvesting complexes~\cite{Bode2009} to quantum dots~\cite{Nirmal1995,Efros1996,Kurtze2012}, where dark excitons form an essential building block for generation of on-demand entangled photon cluster states~\cite{Schwartz2016}.

Here, we demonstrate that a quasi-resonantly driven microcavity polariton condensate is a sensitive probe for the presence of dark excitons and vice versa dark excitons can be utilized to introduce long-lived potentials for a polariton system. Microcavity exciton-polaritons are composite quasiparticles resulting from the strong coupling of photons and bright excitons in a microcavity structure containing embedded quantum wells. They are known to exhibit several kinds of bistability \cite{Bajoni2008,Amthor2015,Pickup2018,Kyriienko2014} or multistability\cite{Gippius2007,Lien2015}, most prominently in the transmission curve when probed quasi-resonantly at an energy slightly above the lower polariton branch using a narrow cw laser~\cite{Baas2004}. We first realize this kind of polariton bistability using the following setup: The sample is a planar GaAs $\lambda$ cavity consisting of 26 top and 30 bottom GaAs/AlAs distributed Bragg reflector layer pairs, containing six In$_{0.1}$Ga$_{0.9}$As quantum wells placed at the central antinodes of the confined light field. The sample shows a Rabi splitting of about 6\,meV and is mounted on the ring-shaped cold finger of a continuous flow helium cryostat at a temperature of 14.8\,K. The measurements are performed at a positive detuning of 1.8\,meV between the cavity and the exciton mode. The linearly polarized cw probe beam is provided by a M-Squared SolsTis cw Ti:sapphire laser  with a line width below 100\,kHz. The laser beam is focused to a spot diameter of about 40\,$\mu$m onto the sample at normal incidence at a detuning of 650\,$\mu$eV with respect to the empty cavity lower polariton mode, which shows a line width of about 170\,$\mu$eV. The light transmitted through the cavity is detected using a 400\,MHz bandwidth photodiode.

\begin{figure}[th]
\centering
\includegraphics[width=1\linewidth]{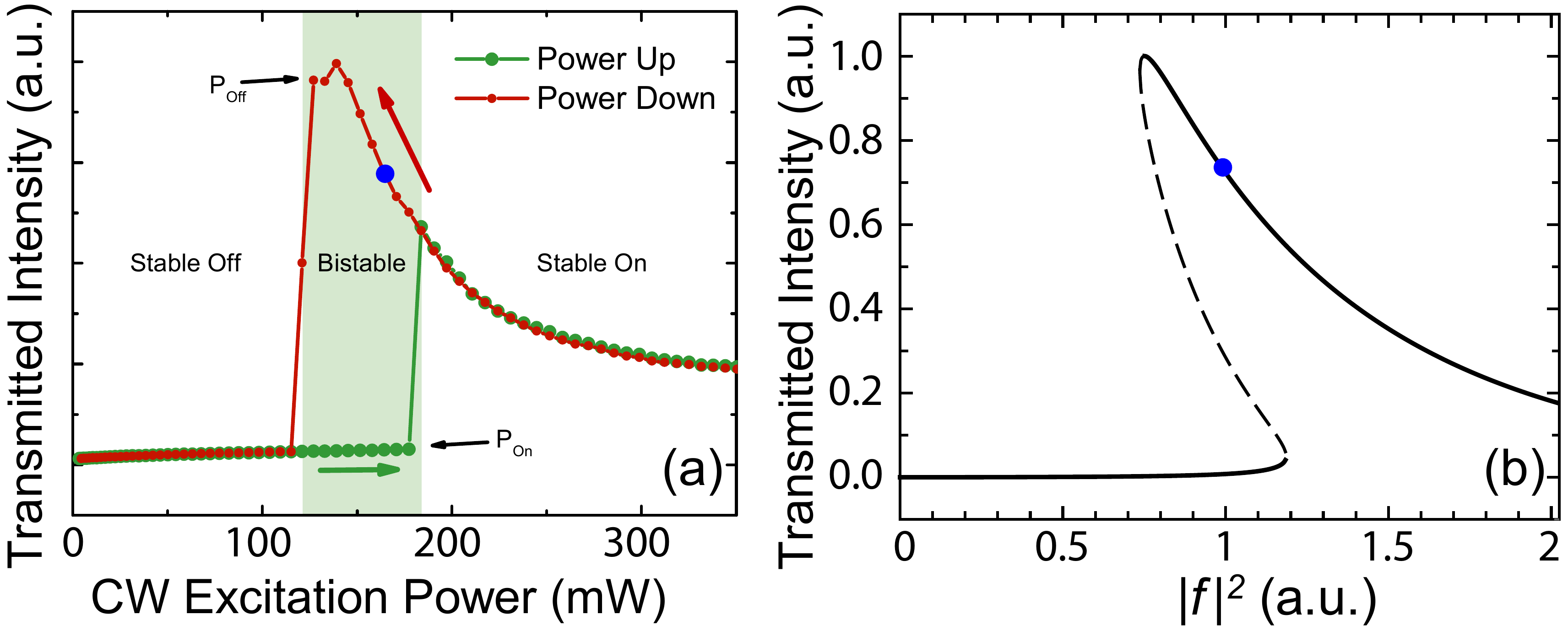}
\caption{
(a) Measured transmission intensity through the sample as a function of the cw excitation pump power.
(b) Transmission through the structure simulated using Eqs.~\eqref{GPEGeneral}--\eqref{XEq}.
Blue circles denote the working position for further discussion.
}
\label{graph:hysteresis}
\end{figure}

Figure \ref{graph:hysteresis}(a) demonstrates the measured hysteresis cycle of the transmission through the sample showing stable off- and on-states and a bistable region in between, which is a consequence of the repulsive interaction of polaritons with the same spin~\cite{Vladimirova2010}. Accordingly, the lower polariton mode experiences a spectral blueshift that depends on the polariton occupation number. Thus, it is the spectral overlap between the lower polariton mode and the probe beam that governs the transmission of the latter through the cavity. The presence of other carriers will also introduce a shift of the polariton mode~\cite{Ouellet2017,Wouters2013,Schmutzler2014b,Menard2014}. As this shift directly translates to a modified probe beam transmission, the latter becomes a sensitive tool to detect the presence of other carriers and measure the strength of their interactions.

Next, we introduce additional carriers into the system and monitor their dynamics using the setup just presented. To this end, we employ a pulsed Ti:Sapphire laser with a pulse repetition rate of 75.39\,MHz and a pulse duration of about 100\,fs to perform far off-resonant excitation at the center of the fourth Bragg minimum of the microcavity structure at 737\,nm. The off-resonant pump is focused to the same sample position as the probe laser, but has a larger diameter of 75\,$\mu$m to ensure that the probe laser samples only the central region of the pump spot. It should be noted that the sample does not show spontaneous condensation under non-resonant excitation. A transition into the weak coupling regime will occur at some point, but all pump powers used here are still below the threshold density for this transition~\cite{SOM}.

\begin{figure}[ht!]
\includegraphics[width=0.95\columnwidth]{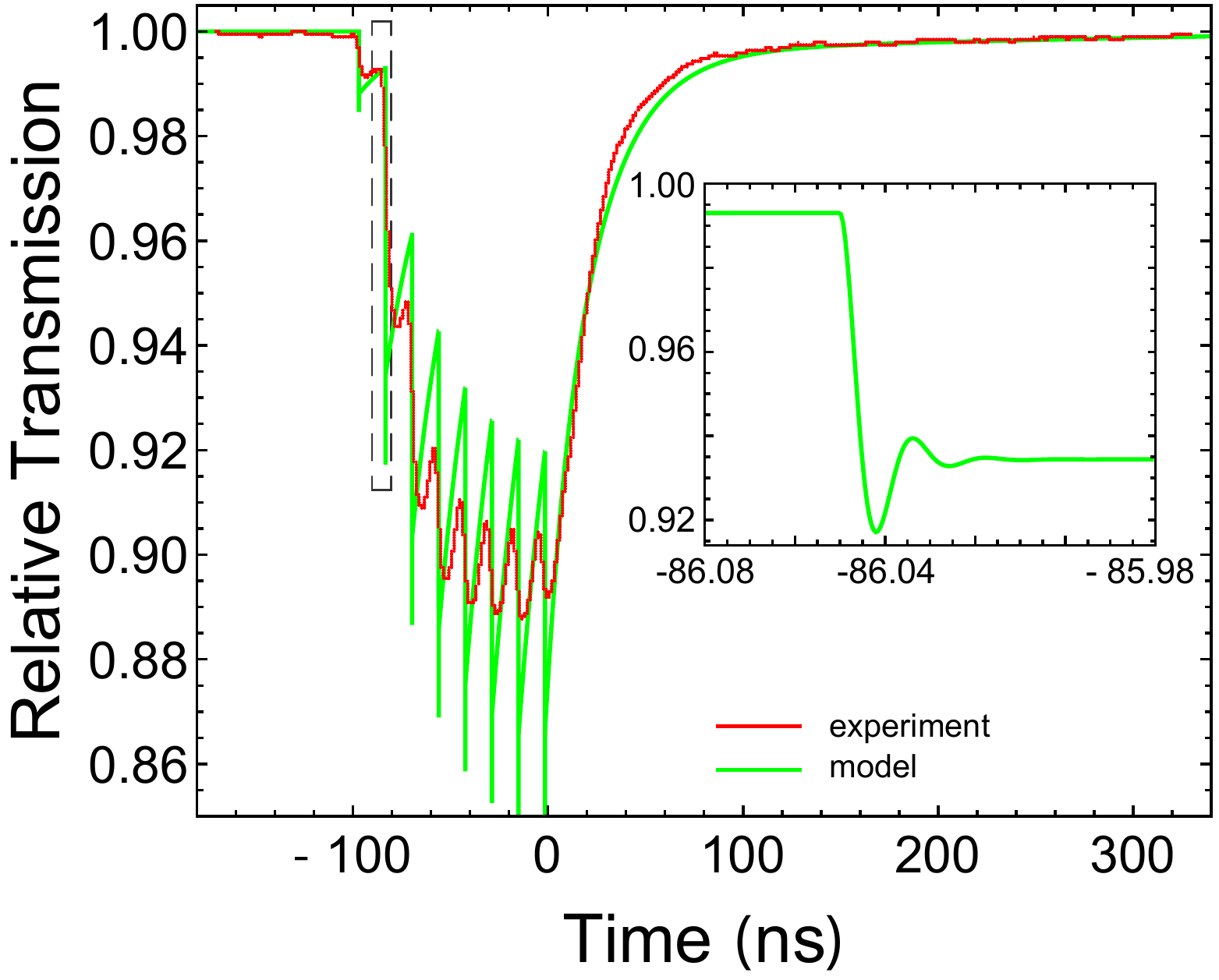}
\caption{
Relative transmission intensity resolved in time. The gate consists of $n = 7$ pulses.
Red and green curves correspond to the experiment and model, respectively.
The slow component shows an exponential decay on a timescale of 22\,ns, while the inset shows fast polariton dynamics on the picosecond scale corresponding to the schematic dashed frame in the main figure.
The cw pump power corresponds to blue dots in Fig.~\ref{graph:hysteresis}.
}
\label{graph:RT_time_dependency_experiment_model}
\end{figure}

In order to investigate time scales longer than the temporal separation between two pulses, we use an electro-optical modulator to gate the non-resonant pump beam. The gate operates at a repetition rate of 100\,kHz and opens for 90 or 103\,ns, which creates pulse trains of seven or eight full consecutive pulses. We set the intensity of the probe laser to an intensity in the middle of the upper branch of the bistability curve as indicated by the blue dot in Fig. \ref{graph:hysteresis}(a) and record the time-resolved change of its relative transmission with respect to the non-resonant pump pulses. Figure \ref{graph:RT_time_dependency_experiment_model} shows a typical trace of the relative transmission. Shortly after a pulse arrives on the sample, the probe transmission diminishes significantly and slowly increases again afterwards. Surprisingly, we find that the relative transmission does not fully recover until the next pulse arrives. Instead the suppression builds up quickly over the course of the first four pulses. Afterwards the peak suppression continues to increase slowly with every additional pulse.
After the last pulse of the train has arrived on the sample, the transmission slowly recovers back to the initial value on a long timescale of tens of ns.

As the reduced transmission translates to a spectral shift of the lower polariton mode, these results raise questions about the nature of the carriers causing this shift. While there have been numerous studies on the dynamics of polariton condensates after non-resonant excitation, the focus has so far been on bright excitations. Free carriers may relax and form bright exciton-like polaritons at large wavevector, which in turn relax down the polariton dispersion via spontaneous or stimulated scattering until they reach the ground state and join the condensate. Both the changes in population dynamics and the presence of free carriers will result in changes of the relative transmission, but they will do so on the short timescale required for carriers to form polaritons, reach the ground state and leave the cavity. Both are typically on the order of tens of ps~\cite{Deng2010}. Even considering a possible slow-down of relaxation at small carrier densities, an upper limit for this timescale is given by the bright exciton lifetime. For high-quality quantum wells, it may be as short as tens of ps, but even for low-quality structures, it will usually not exceed the bulk value of about 1~ns~\cite{Deveaud1991}. Therefore, bright carriers fail to explain the long timescale seen in the experiment. This suggests that optically dark excitations play a significant role for long times after non-resonant excitation.

We reproduce the full set of experimental observations with use of a rate equation model accounting for the long living reservoir of dark excitons that creates a repulsive potential responsible for the blue shift of the polariton condensate energy.
We apply this model to the regime of quasi-resonant cw optical excitation where the bistability curve shown in Fig.~\ref{graph:hysteresis}(a) has been measured as well as to the regime of pulsed excitation where the memory effect has been detected, as Fig.~\ref{graph:RT_time_dependency_experiment_model} shows.
We model the dynamics of the system by solving the Gross-Pitaevskii equations for the polariton condensate wave function $\Psi$ coupled to the rate equations for the occupation numbers of the incoherent reservoir of optically inactive excitons,~$N_{\text{X}}$:
\begin{eqnarray}
\label{GPEGeneral}
 \mathrm{i} \hbar  d_t \Psi &&= \left[ -\delta_{\text{p}} + V_{\mathrm{b}} (t) - \left.  \mathrm{i} \hbar \gamma \right/ 2    \right] \Psi +  f , \\
\label{XEq}
d_t N_{\text{X}} && = P(t) + \beta |f|^2 -  \gamma _{\text{X}}  N_{\text{X}}.
\end{eqnarray}
In~\eqref{GPEGeneral} $\delta _{\text{p}}$ is responsible for the detuning of the resonant pump energy from the bare lower polariton energy, which we choose as a reference.
$V_{\mathrm{b}} (t) = g |\Psi|^2 + g_{\text{X}} N_{\text{X}}$ describes the blueshift of the polariton energy due to the intra-condensate polariton interactions and the interaction with the reservoir excitons;
$g$ and  $g_{\text{X}}$ are the corresponding interaction constants. 
$f$ is the amplitude of the resonant cw pump.
$\gamma$ is the polariton relaxation rate.
Equation~\eqref{XEq} is the rate equation for inactive dark reservoir excitons.
To take into account filling of the reservoir under the resonant pumping we introduce the term~$\beta |f|^2$.
$\beta$ is the dimensional reservoir response constant.
The reservoir is also pumped incoherently by the modulated in time optical pump~$P(t)$.
The exciton reservoir relaxes at a rate of~$\gamma _{\text{X}}$.

Under solely resonant pumping, when one assumes $P=0$, the system has been extensively considered for bistability and related effects.~\cite{Baas2004,Wouters2007,Gavrilov2014, Cancellieri2014}.
Following~\cite{Gavrilov2014}, within the one-mode approximation, $\Psi = \psi_{\mathrm{p}} e ^{-i E_{\text{p}} t / \hbar}$, for the driven cavity polariton mode $\psi_{\mathrm{p}}$ we obtain:
\begin{equation}
\label{EqForDrivenMode}
|\psi _{\mathrm{p}}|^2 = \left. |f|^2 \right/ \theta,
\end{equation}
where $\theta = {(\delta _{\mathrm{p}} - g |\psi _{\mathrm{p}}|^2 - g_{\text{X}} \beta |f|^2 / \gamma_{\text{X}})^2 + (\left. \hbar \gamma \right/2)^2}$.
The calculated transmission intensity through the structure is given by
\begin{equation}
T \propto \left.|\psi _{\text{p}}|^2 \right/ \theta,
\end{equation}
Figure~\ref{graph:hysteresis}(b) shows the transmission $T$ as a function of the cw resonant pump power $|f|^2$.
The parameters used for modelling are given in Ref.~\cite{ParamForMod}.
Two branches (solid) corresponding to stable solutions of Eq.~\eqref{EqForDrivenMode} nicely qualitatively reproduce the experimental dependence for the transmission shown in Fig.~\ref{graph:hysteresis}(a).
The decay in the transmission intensity of the upper hysteresis branch is due to the blueshift of the cavity polariton energy from the pump energy.
The blueshift is caused by polariton interactions with the dark exciton reservoir, which may be populated even in the presence of only the resonant pumping in the positive detuning regime.
This model is aimed at capturing the essential role of dark excitons in cw and pulsed transmission experiments. It deliberately neglects various additional effects such as spin-anisotropic interactions, cavity anisotropies, scattering from the condensate towards the reservoir and non-linear loss due to biexciton formation~\cite{Wouters2013,PhysRevLett119097403,JETPLetters921712010}.

To model the transmission dynamics, we solve Eqs.~\eqref{GPEGeneral}--\eqref{XEq} numerically in the presence of the non-resonant optical gate.
We take the latter as a train of sub-picosecond Gaussian pulses in the form ${P(t) = \sum_{j=0} ^{n-1}} P_{0} \exp \left[ -(t - j/\nu- t_0)^2/w^2 \right]$, where
$n$ is the number of pulses in one train,
$\nu$ is the pulse repetition rate in one train,
$t_0$ is the time of arrival of the first pulse peak,
$w$ is a single pulse duration.
The green curve in Fig.~\ref{graph:RT_time_dependency_experiment_model} shows the transmission variation in time in presence of the optical gate of $n=7$ pulses.
To take into account non-instantaneous opening of the gate, we assume that an additional pulse enters the system prior to the main train.
The pulse possesses an energy of one tenth of the energy of subsequent pulses.
The simulated slow dynamics at the nanosecond scale fully reproduces the measurements.
The inset in Fig.~\ref{graph:RT_time_dependency_experiment_model} shows fast dynamics on the scale of tens of picoseconds.
It reflects the population relaxation after the pulse arrival, see also~\cite{NatureCommun420082013}. The monotonic region after arrival of the last pulse in Fig.~\ref{graph:RT_time_dependency_experiment_model} allows us to estimate the lifetime of dark excitons as $1/\gamma _{\mathrm{X}} \approx 22$~ns, see Supplementary material~\cite{SOM} for the details of the estimation.
Based on the simulations in Fig.~\ref{graph:RT_time_dependency_experiment_model}, we are able to estimate the blueshift provided by the train of seven pulses of a given energy as about 40~$\mu\text{eV}$ achieved at the dark exciton density of about~$5 \cdot 10^8 \, \text{cm}^{-2}$.

One can see that both in cw and pulsed excitation cases the model captures the essential manifestations of the dark exciton reservoir.
Namely, in Fig.~\ref{graph:hysteresis}(b) we reproduce the characteristic decrease of the transmission signal as a function of the pump power that is a signature of the detuning of the condensate energy from the laser mode energy that is governed by population of the dark reservoir.
In Fig.~\ref{graph:RT_time_dependency_experiment_model} the model quantitatively reproduces the dependence of the transmission modulation induced by laser pulses on the reservoir density created by previous pulses.

Several types of excitations could be at the heart of the long-lived line shifts. Parity-forbidden and spatially indirect excitons are unlikely candidates. In addition, coherent multidimensional spectroscopy has demonstrated that they usually show some weak coupling to bright states, which limits their lifetime drastically \cite{Tollerud2016}. The same holds true for the nominally dark $J_z=\pm1$ antisymmetric polariton states that form in microcavity structures containing more than one quantum well. Due to coupling with leaky modes, their lifetime is reduced drastically to values below 1\,ns \cite{Richard2005}. These states form a possible decay channel for dark states, but as they are delocalized, the overlap integral between dark excitons and these states is expected to be small. For biexcitons, also much shorter lifetimes are expected \cite{Wouters2013}. Two kinds of dark excitations should be retained as candidates for the observed dark carrier population. First, spin-forbidden dark excitons with an exciton spin projection of $J_z=\pm2$ may form under non-resonant excitation. These excitations can only decay non-radiatively or by spin relaxation towards a bright state. Second, spin-allowed carriers with $J_z=\pm1$ may form at large wavevectors $k_{||}$. If their wavevector exceeds that of light inside the medium, they also cannot couple to light fields and are thus optically dark. A closer look at the typical relaxation processes already sheds some light on the processes taking place.

In quantum wells not embedded inside a microcavity, the spin relaxation time between dark $J_z=\pm2$ excitons and bright $J_z=\pm1$ excitons is governed by the short range exchange interaction between excitons ~\cite{Snoke1997}, which causes an energy splitting of about 80~$\mu$eV between the bright and dark states with dark states being at lower energy~\cite{Vina1999}. This splitting results in a spin relaxation timescale of about 80\,ps. For quantum wells embedded into a microcavity, the situation changes drastically. In the strong coupling regime, the light-matter interaction shifts the bright state to lower energies by a value given by half the Rabi energy. As this splitting is significantly larger than the splitting in bare quantum wells, also the spin relaxation time by exciton-exciton interaction is expected to become much longer at small $k_{||}$. However, as the splitting depends on $k_{||}$ and due to symmetry reasons, mixing of bright and dark states occurs at $k_{||}\neq 0$ \cite{Gautham2017}, especially in the bottleneck region~\cite{Shelykh2005}. Therefore, it is expected that primarily dark excitons with $J_z=\pm2$ at $k_{||}=0$ will show a drastically enhanced lifetime. Due to the large value of the Rabi splitting, it is expected that relaxation will mostly occur via phonons to bright polariton states in the bottleneck region of the dispersion or with some small probability towards the antisymmetric dark polariton states. Both processes will not depend strongly on the dark exciton density. For $J_z=\pm1$ excitons at large wavevector, momentum and energy relaxation towards the optically active region is supposed to be the most important relaxation channel. Thus, exciton-exciton scattering should play a significant role and some kind of density dependence is expected.

\begin{figure}[ht]
\includegraphics[width=0.95\columnwidth]{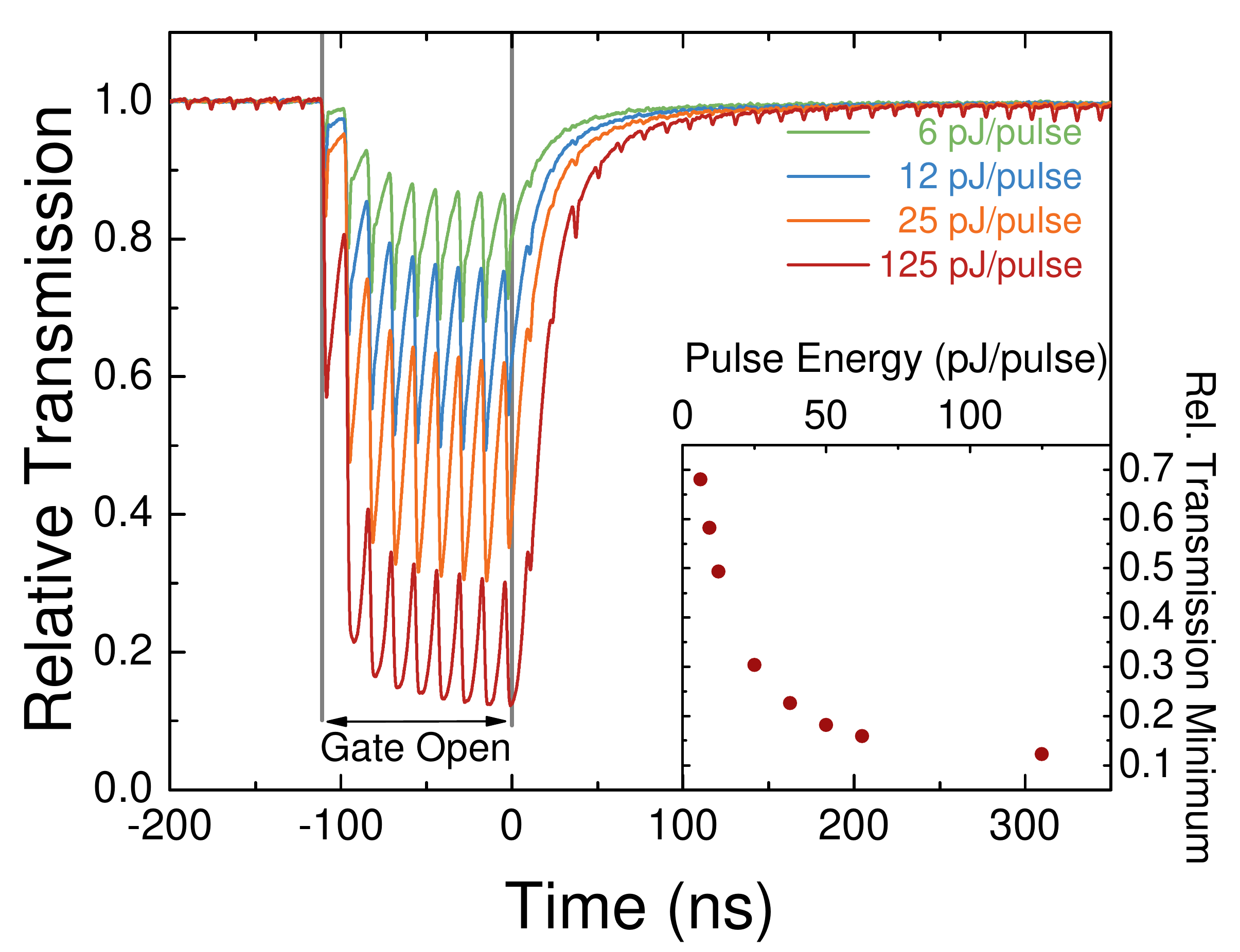}
\caption{The relative transmission intensity resolved in time after the arrival of a train of 8 pulses for different pumping energies. The dashed lines represent the temporal region where the gate is open and the system is excited non-resonantly with a pulse repetition rate of $\nu = 75$\,MHz. The periodic signal at large pump energies arises due to the finite extinction ratio of about 1:100 of the intensity modulator. Inset: Circles represent the minimum value of the relative transmission for different pumping energies.}
\label{graph:transmission_control_various_energies}
\end{figure}

In order to gain some insight on these scenarios and also to estimate the magnitude of suppression of transmission we are able to achieve, we compared the dynamics of the suppression for different non-resonant pump powers as shown in figure \ref{graph:transmission_control_various_energies}. First, indeed the suppression can be enhanced by pumping more strongly. The transmission can be reduced to values below 15\,\% of its initial value. Second, there is no apparent dependence of the relaxation timescale on the non-resonant pump intensity. Accordingly, although the microscopic nature of the dark carriers in our experiment is not known unambiguously, we cautiously suggest that spin-forbidden dark excitons at low momentum are the most likely candidates. Additionally, we also found compelling evidence that the interaction between them and bright polaritons is repulsive: When driven below the non-linear threshold, additional pulsed non-resonant excitation significantly enhances the transmission, which is a signature of an interaction-induced blueshift of the polariton mode~\cite{SOM}.

In summary, we have demonstrated that a narrow polariton mode may be utilized as a sensitive probe for the presence of dark excitations in a semiconductor system. We found that these carriers have a surprisingly long lifetime of more than 20~ns. Besides the possibility to unveil the dynamics of optically dark excitations, which are difficult to address otherwise, our result has several important implications. First, it demonstrates the possibility to optically imprint potential landscapes for polaritons that last three orders of magnitude longer than the polariton lifetime in the system. This provides interesting perspectives for functional polariton circuits and classical polariton simulators~\cite{Berloff2017,Tosi2012,Assmann2012,Askitopoulos2018,Schmutzler2015}. Resonant injection of dark excitons via two photon absorption~\cite{Gautham2017,Lemenager2014,Schmutzler2014} might provide means to create tailored optical potentials without perturbing relaxation dynamics. Finally, typical pulsed excitation experiments on polariton systems employ lasers with a pulse separation of about 13~ns. The existence of dark excitations with a lifetime longer than that implies that the standard assumption that the system is completely empty before an excitation pulse arrives is not tenable, which is of high importance for studies of condensate formation.

\acknowledgements
We gratefully acknowledge support from the DFG in the framework of TRR 160 within project B7.
ESS acknowledges support from RFBR Grants No. 16-32-60104 and No. 17-52-10006.
AK acknowledges financial support from St-Petersburg State University within research grant 11.34.2.2012 and partial support from the Royal Society International Exchange Grant No. IEC/R2/170227.
CS acknowledges support from the DFG in the framework of project~SCHN1376/3-1.


\begin{thebibliography}{44}
\bibitem [{Par()}]{ParamForMod}%
  \BibitemOpen
  \href@noop {} {}\bibinfo {note} {The polariton energy detuning is $\delta
  _{\text{p}} = 650 \, \mu \text{eV}$. The interaction coefficient is $g =
  g_{\text{X}}/3 = 0.6 \, \mu \text{eV}$. The polariton relaxation rate is
  $\gamma = 0.3 \, \text{ps}^{-1}$. The pulse repetition rate is $\nu ^{-1} =
  13.3$~ns, duration of one pulse is $w = 200$~fs.}\BibitemShut {Stop}%
\end{thebibliography}
%

\section{Supplementary Information}
\begin{figure}[tbh]
\includegraphics[width=1\linewidth]{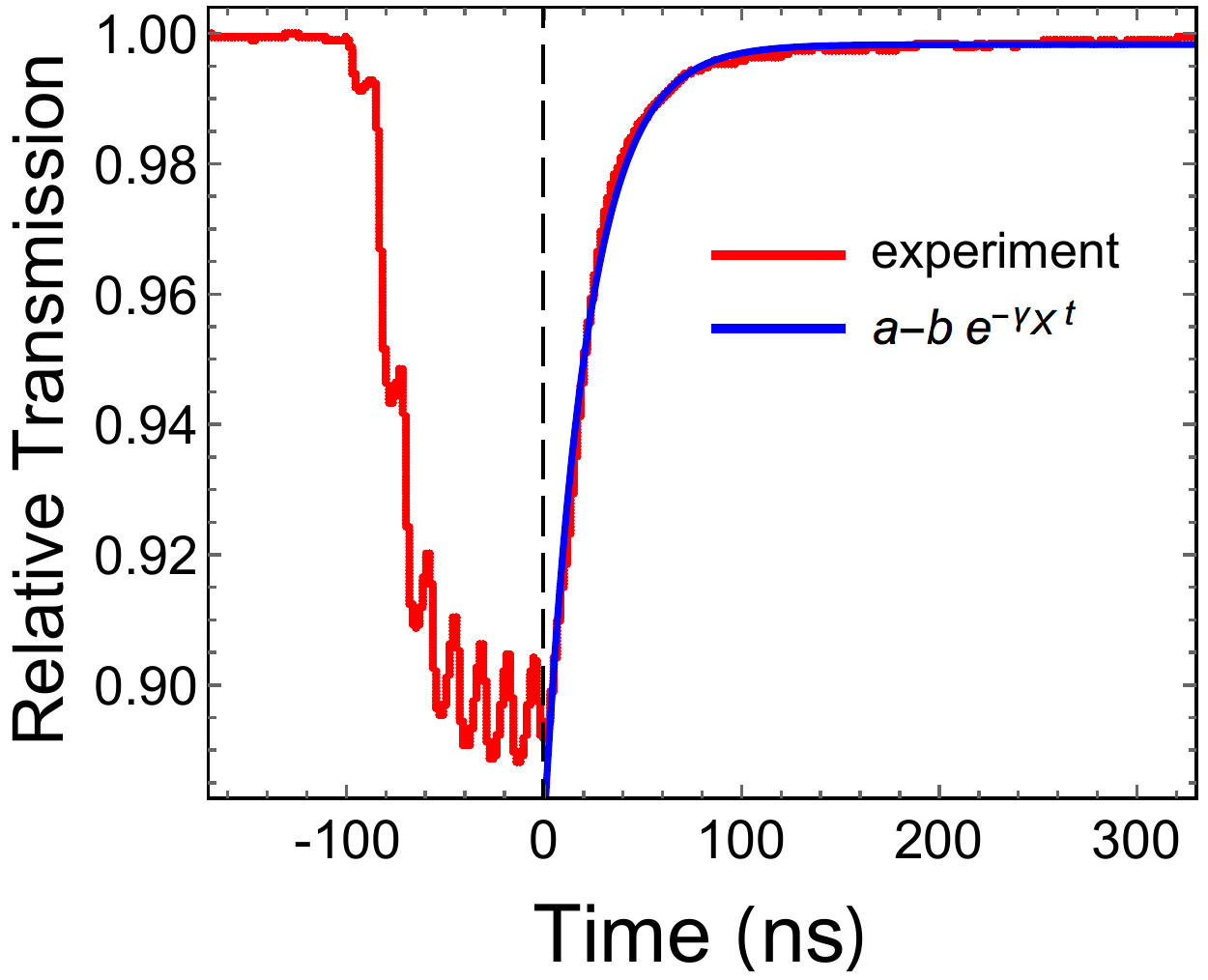}
\caption{ \label{FIGSMXLifetime}
Relative transmission in time.
Red curve corresponds to the experiment.
Blue curve shows fitting of the transmission after the last pulse, $t>0$. 
The fitting parameters are $a \approx 0.998$, $b \approx 0.12$ and $ \gamma _{\mathrm{X}} \approx 45.16 \, \mu\mathrm{s}^{-1}$.
}
\end{figure}

\section{Estimation of the lifetime of dark excitons}

Figure~\ref{FIGSMXLifetime} and Fig.~2 in the main text show the measured relative transmission variation in time.
The dependence consists of three regions: before the gate opens (before $\approx -100$~ps), while the gate is open (from $-100$ to $0$~ps) and after the gate closes (after~$0$~ps).
The third region allows one to estimate the lifetime of dark excitons forming the gate.

Since the duration of each pulse in the train (less than 200~fs) is significantly shorter than other characteristic time scales in the system including lifetimes of polaritons and especially dark excitons, one can assume that each non-resonant pulse instantaneously pumps the reservoir of dark excitons, followed by its slow relaxation until the subsequent pulse arrival. 
After the last pulse of the train enters the cavity, the occupation number of the dark exciton reservoir starts to decrease monotonically until the end of the system evolution.
Rate equation (2) in the main text has a simple solution in this region:
\begin{equation}
\label{XReservSolution}
N_{\mathrm{X}} = N_{\mathrm{X0}} e^{-\gamma _{\text{X}} t} \quad \text{for} \, t>0,
\end{equation}
where $N_{\mathrm{X0}} = N_{\mathrm{X}} (t=0) $. 

Using the fact that the polariton lifetime is much (at least three orders) shorter than that of dark excitons, we can assume the occupancy of the polariton state as well as the transmission through the structure adiabatically adjust to the variation of the occupancy of the dark exciton reservoir.
We fit the relative transmission $T$ for $t>0$ by the function $a-b \exp[-\gamma _{\mathrm{X}} t]$, see Fig.~\ref{FIGSMXLifetime}.
The best matching has been achieved for the fitting parameters $a \approx 0.998$ and $b \approx 0.12$, that allowed us to estimate the relaxation rate and lifetime of dark excitons as $ \gamma _{\mathrm{X}} \approx 45.16 \, \mu\mathrm{s}^{-1}$ and $\tau _{\mathrm{X}} = \left. 1 \right/ \gamma _{\mathrm{X}}  \approx  22.15 $~ns, respectively.

\section{Extended theoretical model}
\begin{figure}[ht!]
\includegraphics[width=0.95\columnwidth]{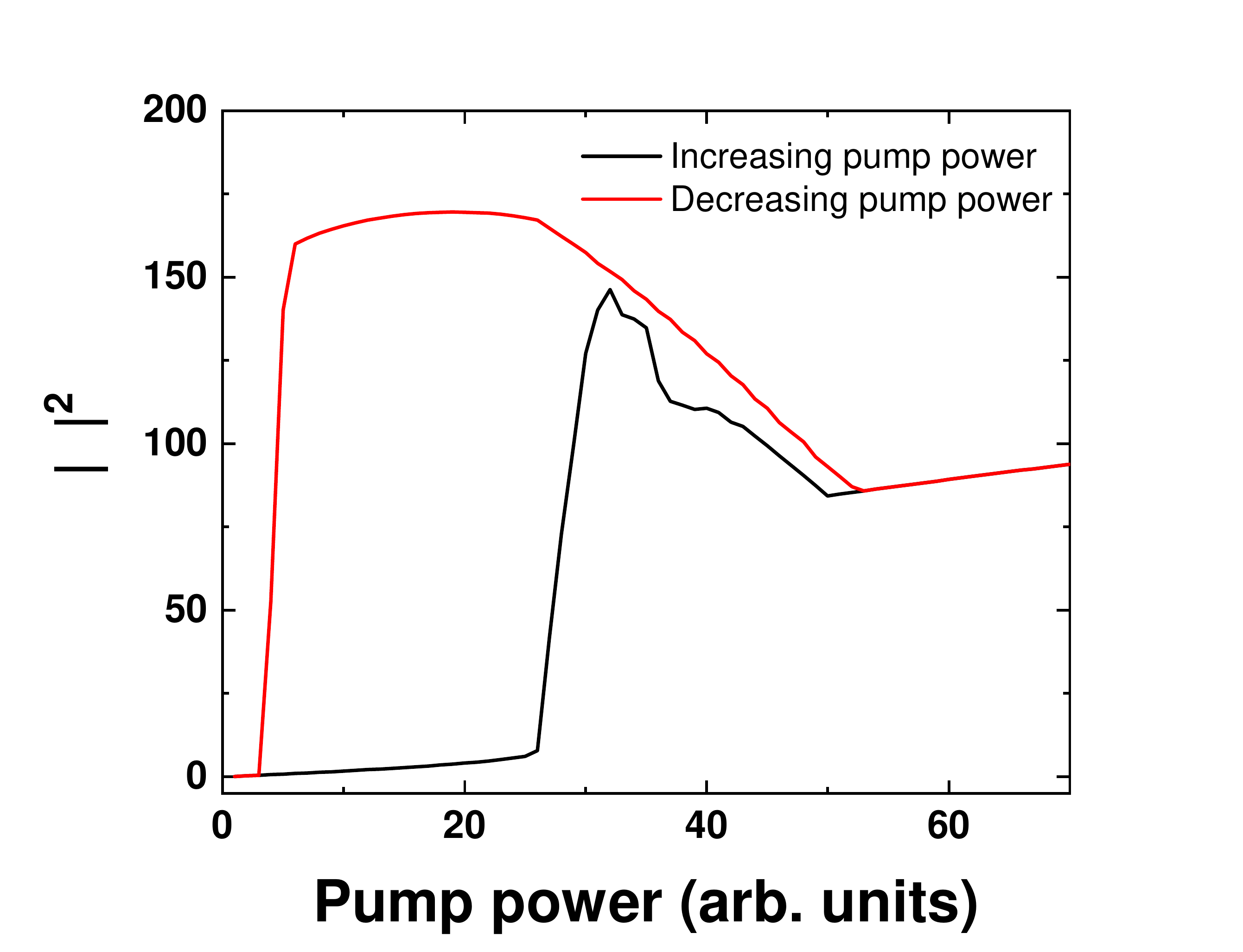}
\caption{Photonic part of the polariton wavefunction in the presence of an inhomogeneous strain distribution.}
\label{graph:IOStrain}
\end{figure}

\begin{figure}[ht!]
\includegraphics[width=0.75\columnwidth]{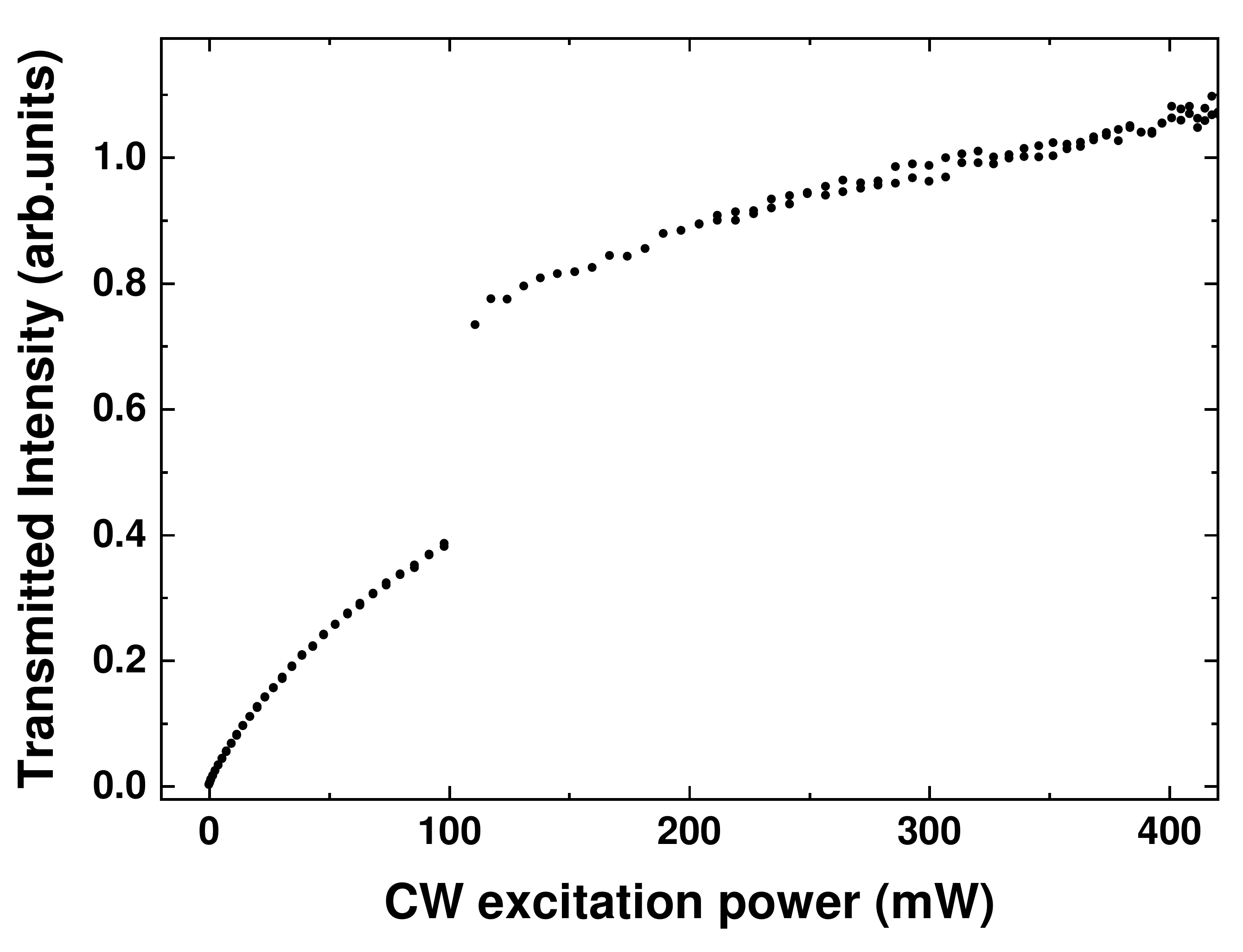}
\caption{\label{FigNoHeating} Transmitted intensity for a slightly positive detuning of the pump beam of less than one linewidth. No heating-related reduction of the transmitted intensity at high powers is observed.}
\end{figure}

A full treatment of the transmission through a resonantly driven microcavity is a highly non-trivial problem. However, in this manuscript we are solely interested in transients after perturbation of the polariton system. We also focus on long-term dynamics that take place at least 1\,ns after the perturbation. For this reason, the main text contains a model that focuses only on these aspects and treats all other influences on a phenomenological level. It is drastically simplified, but also much easier to understand and only treats the relevant effects explicitly. In the following we discuss the full model, the approximations we applied and the physical origin of the shape of the transmission curve. The model follows the one presented in \cite{Wouters2013}, where also an in-depth discussion of polariton bistability and multistability for non-circularly polarized pumping can be found.\\
In polariton bistability experiments, the spectral shifts of the polariton line due to interactions effectively change the detuning between the cavity and the exciton resonance. A full treatment of these effects requires working in the exciton-photon basis with coupled equations for the photon and exciton fields $\phi$ and $\chi$ that form the polariton. Also, polariton interactions are spin-dependent, so a full treatment has to take into account that for both fields two spin species exist. This results in the following set of 5 equations:
\begin{align}
i\frac{d}{d t}\chi_{\uparrow,\downarrow}=&\frac{\Omega_R}{2}\phi_{\uparrow,\downarrow}+[\Omega_X-\frac{i}{2}(\gamma_X+\beta |\chi_{\downarrow,\uparrow}|^2)\nonumber\\
&+g_R n_R +\alpha_1|\chi_{\uparrow,\downarrow}|^2 +\alpha_2 |\chi_{\downarrow,\uparrow}|^2]\chi_{\uparrow,\downarrow}\\
i\frac{d}{dt}\phi_{\uparrow,\downarrow}=& \frac{\Omega_R}{2}\chi_{\uparrow,\downarrow}+(\Omega_C-\frac{i}{2}\gamma_C)\phi_{\uparrow,\downarrow}\nonumber\\
 &+\frac{\Omega_{\text{lin}}}{2}\phi_{\downarrow,\uparrow} + P_{\uparrow,\downarrow}\\
\frac{d n_R}{d t}=&2\beta |\chi_{\uparrow}|^2|\chi_{\downarrow}|^2 - \gamma_R n_R + P_R
\end{align}
Here, $\Omega_R$ represents the Rabi frequency and couples the photonic and excitonic fields. $\hbar \Omega_X$ and $\hbar \Omega_C$ give the energies of the bare exciton and cavity modes, respectively, where all energies are given relative to the energy of the polarized quasi-resonant cw pump laser, which pumps the photonic modes at rates of $P_{\uparrow \downarrow}$. $n_R$ represents the occupation number of the dark reservoir. It may consist of any of the optically dark excitations mentioned in the main text. It decays with a decay rate $\gamma_R$ and is fed by two terms. For non-resonant pumping free electrons and holes are formed. After relaxation some of them will form dark excitons at a rate $P_R$. $\hbar \Omega_{\text{lin}}$ describes the linear polarization splitting, which effectively couples the two photonic modes. The interaction-induced blueshift of the polariton mode is taken into account via the exciton field. Here, $\alpha_1$ and $\alpha_2$ describe the interaction strength between polaritons of the same or different spin, respectively. Generally speaking, $|\alpha_1|>|\alpha_2|$ and $\alpha_1$ is positive, resulting in a repulsive interaction between polaritons of the same spin, while $\alpha_2$ is negative, which yields an attractive interaction between polaritons of opposite spin. $g_R$ describes the interaction between polaritons and the dark reservoir. As the spin degree of freedom for the dark reservoir is not explicitly taken into account, $g_R=\frac{\alpha_1+\alpha_2}{2}$ represents an averaged interaction strength. Also, a non-linear loss term from the ground state via biexciton formation at a formation rate $\beta$ is included. In this process, two excitons of different spin form biexcitons. This process is considered to be especially efficient at positive detunings, where the polariton energy is close to the biexciton resonance. These biexcitons will in turn directly interact with the polariton condensate or decay and form bright or dark excitons. In the experiment, the bright excitons will be short-lived due to stimulated scattering towards the ground state and biexcitons are supposed to decay also on a sub-ns timescale. For cw steady state pumping, therefore some equilibrium between formation and decay of biexcitons will take place. At this point, we perform the first major approximation. Instead of treating biexcitons, bright excitons and dark excitons seperately, we only consider a long-lived dark reservoir and neglect the short-lived particles. Therefore, for steady state cw pumping in the absence of an additional non-resonant pump, the occupation of the dark reservoir will be overestimated slightly. However, in this regime polariton-polariton interactions dominate the blueshift of the lower polariton mode. Further, there are no non-linear decay channels for the dark reservoir, so this approximation just changes the base number of dark excitons present, but does not influence the transients significantly. For the same reason, also for the pulsed non-resonant pump only dark exciton formation is considered as bright excitons will undergo stimulated scattering towards the condensate quickly.\\
It is well known that for linearly or elliptically polarized pumping, the transmission through the cavity is high when only one of the two spin species is on the upper branch of the bistability curve, while the transmission is actually reduced due to effective population of the reservoir as soon as both spin species reach the upper state \cite{Wouters2013}. The upper bistability threshold therefore depends strongly on the pump polarization and also on the linear polarization splitting of the microcavity. In turn, the latter is heavily influenced by the strain present within the microcavity structure which is necessarily spatially inhomogeneous for samples mounted in transmission geometry. The total transmission through the cavity is then given by the photonic part of the polariton wavefunction averaged over all realizations of the local strain landscape. Figure \ref{graph:IOStrain} shows the effects of the strain landscape of up to 250\,$\mu$eV for pumping with a small circular polarization degree of 0.08.
As can be seen clearly, the strain distribution results in locally varying shifts of the position of the upper bistability threshold and thereby transforms the abrupt reduction in transmission into a smooth decrease. The shape of this transmission curve agrees qualitatively with the one experimentally measured. A more precise agreement may be achieved in principle by measuring the exact strain distribution, which is, however, subject to day-to-day fluctuations during cooling of the sample. As the investigations of transients in the main text are carried out at a single pump power on this curve, the agreement is fully sufficient.\\
At this point, we can separate the fast and slow dynamics and simplify the model to the phenomenological one given in the main manuscript by neglecting terms that only contribute to the fast dynamics. As we work only at one fixed cw pump power, we also replace all terms that are relevant to the shape of the bistability curve, but do not contribute significantly to the transient dynamics, by simple phenomenological terms. To this end, we avoid working in the rather intransparent exciton-photon basis and replace it by the effective polariton mode occupation $\left|\psi_p\right|^2$ and account for the modified Hopfield coefficients by introducing the effective detuning term $\theta$ that links the polariton occupation to the transmission. Along the same lines, we decouple the biexciton formation process that also depends on the Hopfield coefficients from the population of the polariton mode and instead link it directly to the resonant pump intensity. The effects of this term are anyway minimal and a constant value provides a worst case estimate of its effects. We also neglect the spin degree of freedom as we work at a single point of the bistability curve and the non-resonant pump pulse that causes the transient dynamics introduces unpolarized particles into the system on average. After these simplifications we are left with the model equations given in the main text:
\begin{align}
i\hbar \frac{d \psi}{d t}=&[-\delta_p+V_b(t)-i\hbar\frac{\gamma}{2}]\psi+f\\
\frac{d N_R}{d t}=&P(t)+\beta|f|^2-\gamma_R N_R.
\end{align}
Note that in this supplemental material $N_X$ has been renamed $N_R$ to avoid confusion between terms relating to the reservoir and to the excitonic part of the wavefunction. These equations have a minimized amount of free parameters and focus solely on the slow dynamics introduced by dark excitons.\\
Finally, one might also imagine that the presence of the cw beam alone is sufficient to introduce significant heating of the sample. In that case the redshift of the polariton mode introduced by the sample heating might be considered as a potential reason for the reduced transmission seen in the transmission curve at high pump powers. However, such effects should depend solely on the pump beam intensity. For comparison, we also recorded the transmission curve at a pump laser detuning that is too small to support bistability. As can be seen in figure \ref{FigNoHeating}, the transmission through the sample is not reduced at high pump powers, which clearly shows that this reduction is not associated with thermal heating of the sample, but is intimately connected to the presence of bistability.

\section{Strong-coupling to weak-coupling transition}
\begin{figure}[ht!]
\includegraphics[width=0.75\columnwidth]{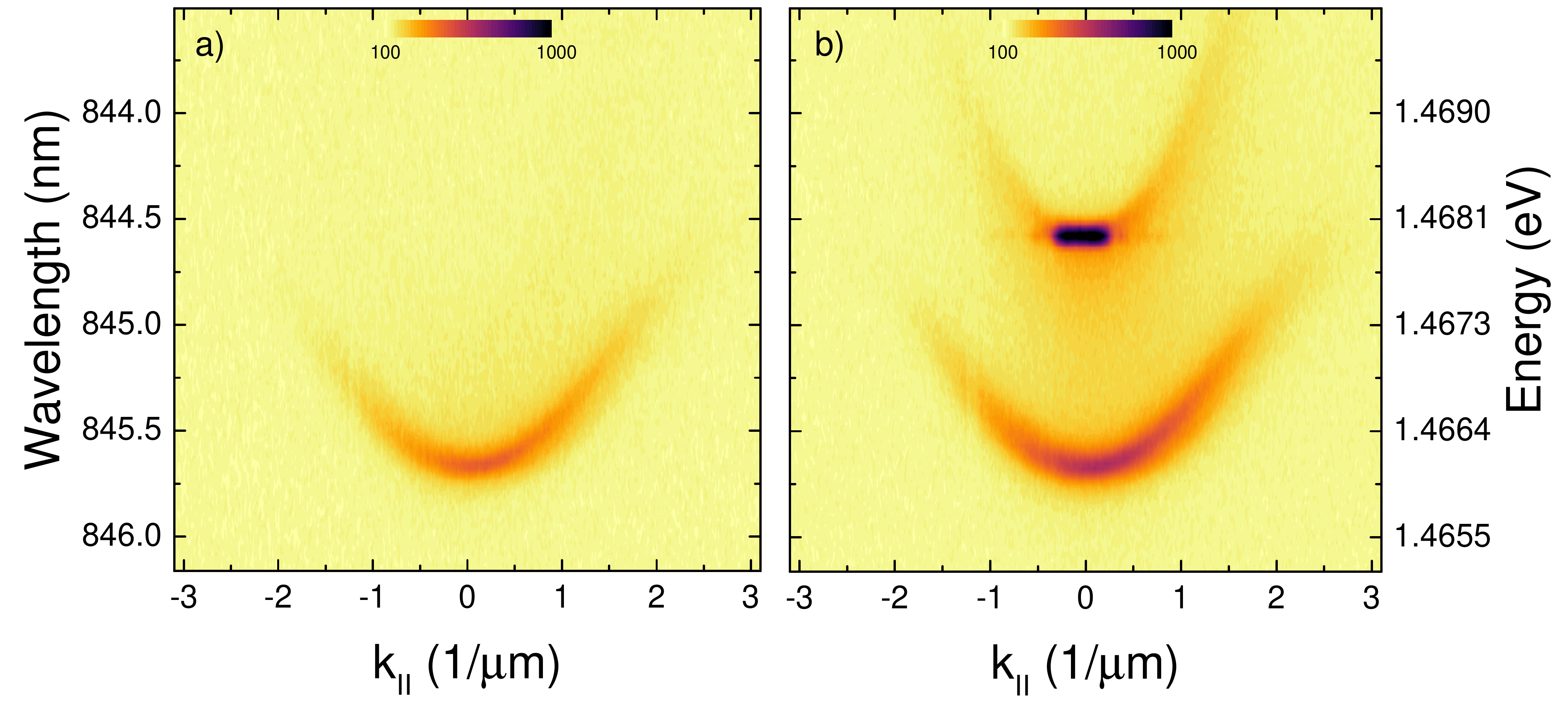}
\caption{Far-field image of the lower polariton branch dispersion for two different non-resonant excitation energies per pulse E$_{\text{NR}}$. a) For E$_{\text{NR}} = 175\,$pJ per pulse, the system is still in the strong-coupling regime, indicated by PL originating solely from the strongly coupled exciton-polariton states. b) For P$_{\text{NR}} = 412.5\,$pJ per pulse, the system is already in the weak-coupling regime, indicated by the additional PL from the cavity mode.}
\label{graph:strong_to_weak_transition}
\end{figure}

\begin{center}
\begin{figure}[ht!]
\includegraphics[width=0.75\columnwidth]{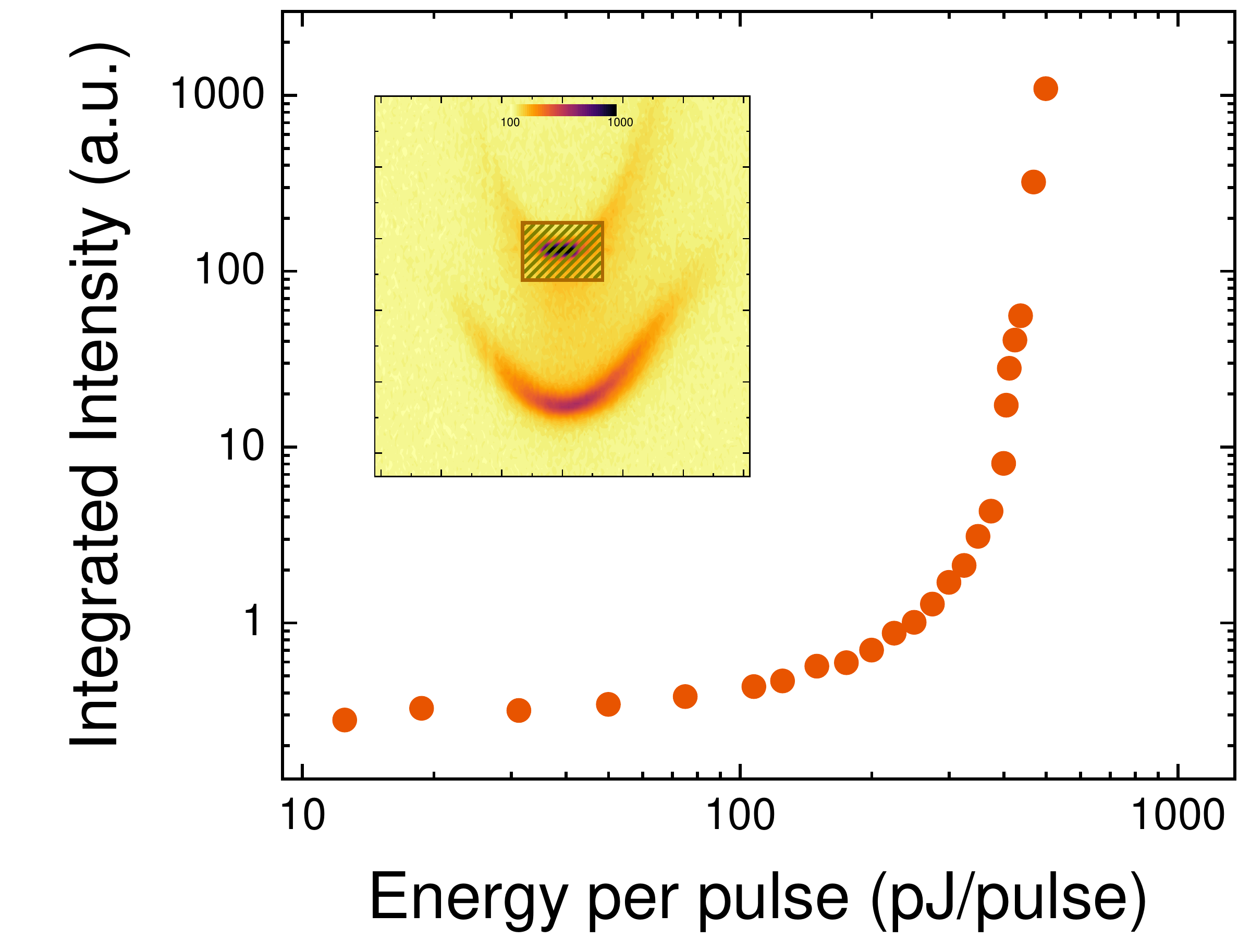}
\caption{Time-integrated output from the bare cavity mode under non-resonant excitation using a train of 8 pulses with repetition rate of $\nu_{\text{pick}} = 100\,$kHz. The vertical axis is the integrated intensity obtained from the brown dashed box in the inset for various energies per pulse. }
\label{graph:strong_to_weak_transition_all_powers}
\end{figure}
\end{center}

When exciting the microcavity sample above the band gap energy, non-resonantly excited carriers may form polaritons and relax towards the lower polariton branch ground state. Whether this process results in spontaneous exciton-polariton condensation depends strongly on the investigated sample. \\
Our sample contains six quantum wells and power dependent measurements with non-resonant excitation show no sign of exciton-polariton condensation as is typical for this sample design. However, with increasing non-resonant excitation powers, we observe a transition from the strong-coupling into the weak-coupling regime. This transition can be identified by additional photoluminescence (PL) at energies above the polariton dispersion, where the PL of the bare cavity mode becomes visible (see Fig. \ref{graph:strong_to_weak_transition}).  
For a more quantitative analysis of the transition from the strong-coupling into the weak-coupling regime, we performed a full set of power dependent measurements. Here, we integrated the PL close to k$_{||} = 0$ for energies around the bare cavity mode. The result as well as the area of integration can be seen in figure \ref{graph:strong_to_weak_transition_all_powers}. A clear non-linear increase of the integrated intensity from the bare cavity becomes visible at approximately 350\,pJ per pulse. This energy per pulse marks the transition point, where the system leaves the strong-coupling regime and therefore loses its polariton eigenstates.
\newpage
\section{Blueshift of polaritons}
\begin{center}
\begin{figure*}[ht]
\includegraphics[width=0.95\textwidth]{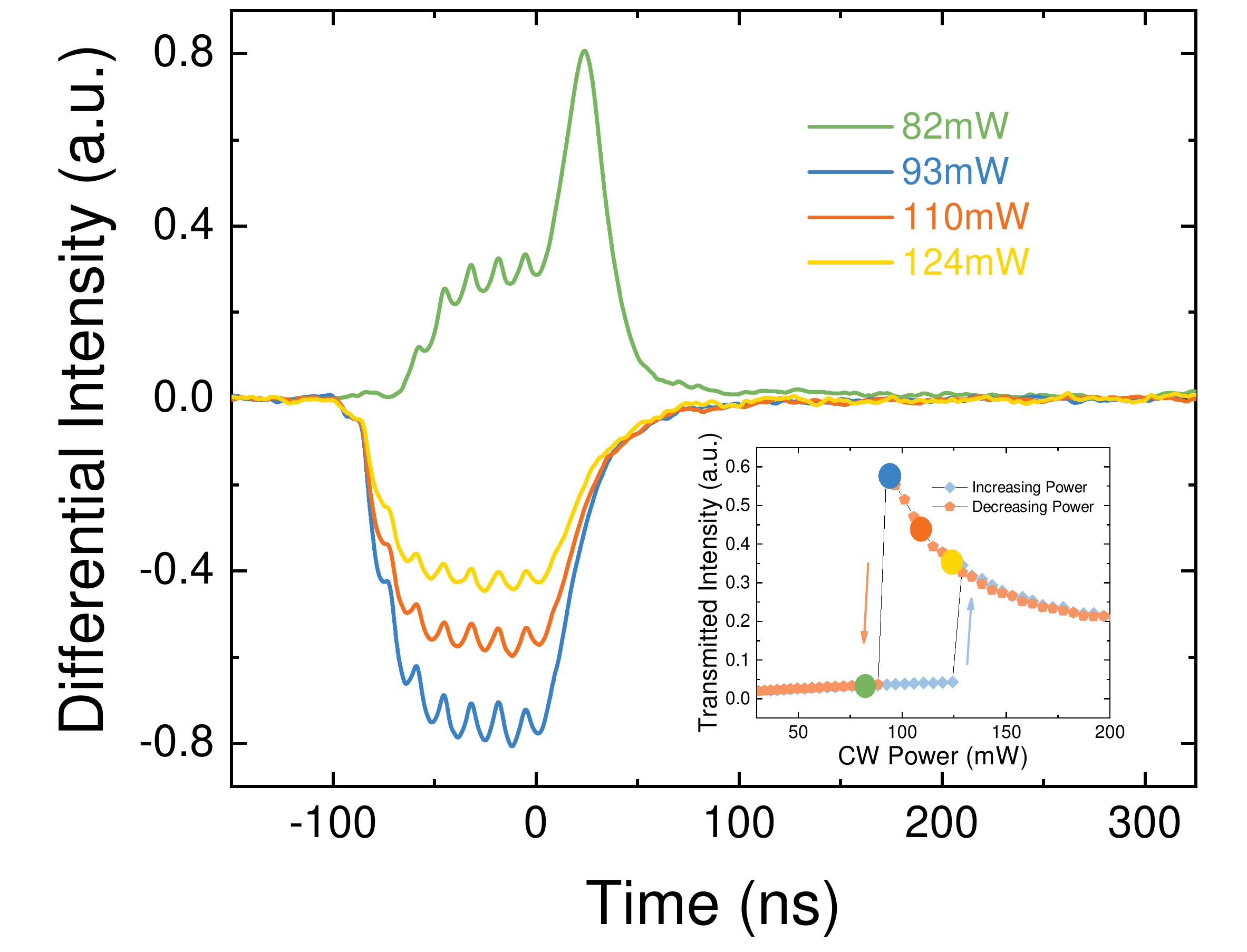}
\caption{Time-resolved transmission measurements for different cw pump powers along the hysteresis curve. The vertical axis shows the differential intensity to create a base line of the intensity for every measurement. Inset: Transmitted intensity of hysteresis curve. The different filled dots in the inset correspond to the cw pump powers chosen for the measurements in the main figure. }
\label{graph:repulsive_interaction}
\end{figure*}
\end{center}

To show that the interaction between the long-lived dark carriers and polaritons is repulsive, we performed time-resolved measurements of the relative transmission for various cw probe beam powers. The powers were chosen such that different significant points along the hysteresis curve are covered (see full dots in the inset of Fig. \ref{graph:repulsive_interaction}). Then, a train of 7 non-resonant pulses is applied to the sample. Interestingly, when the power of the cw probe beam is below the bistable region (green ball in the inset of Fig. \ref{graph:repulsive_interaction}), the response of the system is an overall enhancement of the transmission with an additional more significant enhancement after the end of the pulse train. For these cw probe powers the repulsive nature of the dark carriers quickly blueshifts the polariton dispersion towards and beyond the energy of the cw beam.
After the last pulse of the train has arrived, the dark excitons slowly start to decay again and the polariton mode is shifted towards lower energies. At some point in time, the polariton dispersion becomes resonant with the pump beam again and as energy of the polariton mode changes much slower during the decay of the reservoir as compared to its build-up, the enhancement of the transmitted intensity lasts longer and is more pronounced compared to the initial increase. For all cw pump powers within the bistable region, the transmitted intensity is only reduced. This behavior rules out net attractive interactions between dark excitons and polaritons and also other influences such as sample heating, which would result in a redshift of the polariton mode. Also, the typical time scale associated with the slow decay of the differential intensity is comparable for all cw pump powers within experimental accuracy. This rules out explanations in terms of critical slowing down as in this case, the decay rate should approach the bare polariton lifetime for pump powers below the non-linear threshold \cite{Ballarini2009}.

\end{document}